# Wavelength convertible quantum memory satisfying ultralong photon storage and near perfect retrieval efficiency

Byoung S. Ham

School of Electrical Engineering and Computer Science, Gwangju Institute of Science and Technology
123 Chumdangwagi-ro, Buk-gu, Gwangju 61005, S. Korea
bham@gist.ac.kr
(Submitted on Feb. 20, 2018)

**Abstract:** Quantum coherence control is presented for wavelength convertible quantum memory in a double-lambda-type solid ensemble whose spin states are inhomogeneously broadened. Unlike typical atomic media whose spin decay is homogeneous, a spin inhomogeneously broadened solid ensemble requires a counter-intuitive access in the quantum coherence control to avoid spontaneous emission-caused quantum noises. Such quantum coherence control in a solid ensemble results in a near perfect retrieval efficiency and is applicable to ultralong photon storage up to the spin phase relaxation time. Here, the basic physics of the counter-intuitive quantum coherence control is presented not only for two-photon (Raman) coherent transients, but also for a detailed coherence transfer mechanism resulting in frequency up-/down-conversion. This work sheds light on potential applications of quantum optical memories satisfying noise free, near perfect, ultralong, and multimode photon storage, where quantum repeaters, scalable entangled qubits, and magnetometry would be imminent beneficiaries.

Quantum coherence control in a lambda-type three-level optical ensemble has drawn much attention for various applications of quantum nonlinear optics over the last several decades, ranging from Kerr nonlinearity to quantum information, where a controlled coherence conversion between optical and spin states plays a key role. Of them are electromagnetically induced transparency [1-3], nondegenerate four-wave mixing (NDFWM) [4-11] including phase conjugation [4,5], resonant Raman echoes [6], ultraslow and stopped lights [7-9], stationary lights [10,11], Schrodinger's cats [12,13], entangled photon-pair generations [14,15], off-resonant Raman scattering [16], and photon echo-based quantum memories [17-29]. An essential requirement for the Raman-type nonlinear quantum optics is a long coherence time between two-ground (or spin) states [3]. In the study of ensemble quantum memories such as stopped lights [7-9] and modified photon echoes [17-20], the control of coherence (population) transfer between optical (a common excited) and spin (an auxiliary ground) states plays an essential role for collective phase manipulations of the ensemble. In such a quantum coherence control process, coherence inversion is inevitably induced to the ensemble [17-19], but proper understanding of the system's coherence evolution depending on population change has been limited [9,11,20,23,24]. Although, Kerr nonlinear optics in a pulsed scheme has opened a door to practical quantum optical memories with near perfect retrieval efficiency and ultralong storage time [17], the research on a solid ensemble [9,20] whose spin transitions are inhomogeneously broadened has conflicted with spin homogeneous ensembles [7,8,17] due simply to blind adaptation of the same method applied in a different system. So far, there has been no observation of quantum memory satisfying both near perfect retrieval efficiency and spin coherence-limited storage time, yet.

To overcome the single-mode storage limitation in the early versions of quantum memory such as stopped light [7-9] and off-resonant Raman scattering [16], the intrinsically multimode-capable photon echoes have been studied mostly to solve the population inversion constraint over the last decade [25]. In an inhomogeneously broadened optical ensemble, photon echoes offer the great benefits of Kerr nonlinear optics, such as complete data absorption, efficient echo generation, and storage time extension [17], otherwise the ensemble's inhomogeneity annihilates the system coherence rapidly due to its dephasing process among atoms [26] or spins [20,27]. The multimode storage capability of photon echoes is advantageous for mass information processing compared with cavity quantum electrodynamics [28], off-resonant Raman echoes [16], single color center diamonds [29-31], and even stopped lights [7-11]. Since the first solution model was proposed in 2001 to remove the population inversion constraint in photon echoes by using quantum coherence control in a spin homogeneous Doppler medium [17], several methods have followed in the name of atomic frequency comb (AFC) echoes [21], gradient echoes [22], and controlled double rephasing echoes [18,19]. To lengthen the photon storage time in an optical ensemble, an optical-spin coherence conversion technique in ref. 17, however, has been naively adapted to a non-Doppler solid medium [20] without correct understanding of the ensemble phase shift [18,19], resulting in an absorptive echo [27]. The ultralow retrieval efficiency in photon echo-based quantum memory protocols [32,33] should limit recursive operations such as in circuit-based quantum computing [34] and quantum repeaters for long-distance quantum communications [35]. Short photon storage times have been another drawback of some photon echo-based quantum memories [20-23] especially for quantum repeaters [35]. A long spin coherence time [36-38] is also essential in magnetometry for ultrahigh sensitivity [39,40]. Here, the storage time and the retrieval efficiency in quantum nonlinear optics including photon echoes are inversely proportional to each other. Thus, achieving both long storage time and near perfect retrieval efficiency simultaneously has been a long-lasting goal in quantum optical information processing.

To achieve the long-lasting goal of ultralong photon storage and near perfect retrieval efficiency together, quantum coherence control in conventional nonlinear optics has been adapted to photon echoes for coherence control of ensemble phases, the so-called controlled coherence conversion (CCC) [17-19,27]. Here, traditionally CCC is accomplished simply by adding control Rabi flopping to a two-level (photon echo) system, where CCC couples between the common excited state (atoms, ions, or spins) and an auxiliary ground state [7-11]. A counter-propagating control pulse set for CCC contributes to near perfect retrieval efficiency via a backward echo [17], where the backward echo is free from echo reabsorption in photon echoes [41]. Thus, CCC in solid media has drawn much attention to control the ensemble phase in quantum memories for lengthening the photon storage time and achieving near perfect retrieval efficiency simultaneously [18]. Although the basic physics of CCC has been studied for ensemble phase control [17-19], the solution models for quantum memories are somewhat complicated in such a way of providing a silence echo in a double rephasing scheme [18,19,23,24], preparing thousands of redundant pulses for AFC generation [20,21,32], and tailoring an ensemble spectrum for gradient echoes [22,33]. Moreover, the basic physics of CCC is not well understood in dealing with spin inhomogeneity, and this is the motivation of the present research.

Unlike vastly studied alkali atoms in quantum nonlinear optics [1,4,5,7,8,10,15,16,17,33,36,38], solid media such as rare-earth doped crystals [2,5,6,9,12,20,21,23,24,26,27,32,37,41-44] and color center diamonds [29-31,39,40], have an intrinsic property of spin inhomogeneity. In general, the spin inhomogeneity deteriorates nonlinear efficiency due to ensemble dephasing [20,26,27], so that such a solid ensemble has been strictly prevented from time-delayed operations such as photon storage. However, a spin inhomogeneously broadened solid ensemble has been successfully demonstrated for photon storage by adding a spin rephasing process in both slow light [9] and spin echoes [29]. In such systems, the added spin rephasing does not fulfill the quantum memory requirement of no population inversion, resulting in spontaneous and/or stimulated emission-caused quantum noises [9]. Although the population inversion has no problem at all in a single color center diamond [29] or in accumulated photon echoes [43] such as AFC due to single photon usage for rephasing [20,21,32], the physical mechanism of the ensemble phase shift is not clearly understood yet. Here, the so-called *controlled echo* is proposed, analyzed, and discussed for quantum coherence control in a solid ensemble whose spin transition is inhomogeneously broadened, and a solution model is sought for a *wavelength convertible quantum memory* satisfying both ultralong photon storage and near perfect retrieval efficiency. As a result, general understanding of quantum coherence control in an inhomogeneously broadened ensemble is achieved, where it can be used for various applications of quantum information processing. Incidentally, the importance of solid media for quantum optical information processing is not only in having a general understanding of quantum coherence control, but also is in the practical benefits of no atomic diffusion for ultralong photon storage, including imaging processing, single shot write- and read-outs for fast and reliable processing, qubit scalability in both spectral and spatial domains, and spatial/spectral/temporal multiplexing capabilities for a high dimensional quantum interface.

For the theoretical investigations of the present *controlled echo*-based quantum memory, ensemble coherence evolutions in a spin inhomogeneously broadened medium are studied to determine why the conventional control access should fail and what has misled in physics so far. For this, a simple lambda-type three-level optical system is first investigated to derive the basic physics of quantum coherence control for the *controlled echo* in Figs. 1 and 2. Then, a *wavelength convertible quantum memory* is presented in a double-lambda-type four-level system in Figs. 3 and 4. For both cases, the essential requirement is the stable two-photon (or Raman) coherence between two ground states. The optical inhomogeneous broadening hardly affects the two-photon coherence in such a Raman scheme, where the two-photon coherence plays a key role for the optical-spin coherence conversion and vice versa via CCC [9,17-19]. For this, we speculate the control pulse access-dependent coherence inversion and the resultant population redistribution in the interacting ensemble. The fundamental difference of the present *controlled echo* from the previous photon echo-based quantum memories [17-25] is in the complete exclusion of optical inhomogeneity for rephasing mechanism. It does not support Raman echoes, either [6,33,44]. In that sense, the present scheme roots in conventional quantum nonlinear optics such as slow light-based quantum memories [7-10], but the use of spin inhomogeneity is exclusive and essential for the applications of coherent transients.

Figure 1(a) shows the energy level diagram of a spin inhomogeneously broadened solid ensemble for the study of the basic physics of the present *controlled echo*. The pulse sequence for Fig. 1(a) is given in Fig. 1(b), where the pulse area of the Raman rephasing pulse R is $2\pi$ and composed of two resonant optical pulses A and B [6,9,44]. The control pulse set of B and C together works for CCC of the data pulse A-excited ensemble via Rabi flopping between the excited state $|3\rangle$ and the auxiliary ground state $|2\rangle$ [18,19,27]. Here, the novelty of the *controlled echo* is in the opposite access of the control pulse C, which is counter-intuitive compared to conventional ones [4-14], to compensate for the Raman rephasing-caused population swapping between two ground states. To talk about the control pulse access in the ensemble of Fig. 1, a complete understanding of NDFWM must be preceded for a general scheme of quantum nonlinear optics. So far, no coherence evolution in NDFWM has been thoroughly studied in such a system. On the contrary, the conventional control pulse access has already been demonstrated in atomic [7,8,38] and solid media [6,9,44] for either resonant Raman echoes [6,44] or ultraslow light-based photon storage [7-9,38]. Here, the conventional control pulse access stands for the use of an identical control pulse set (B=C) for the same optical transition, where orthogonal polarizations are normally taken.

The procedure of the *controlled echo* in Fig. 1 is as follows. First, the data pulse A–excited coherence $\rho_{13}$ at $t = t_A$ is the original optical coherence created in the ensemble via a quantum interface between a photon(s) and an ensemble. Initially all atoms are in state $|1\rangle$: $\rho_{11}(t_0) = 1$; $\rho_{ij}(t_0) = 0$. Because the original coherence $\rho_{13}$ by the data pulse A results in the population excitation on the excited state $|3\rangle$ and vice versa, the control of the excited state population $\rho_{33}$ is the heart of the quantum coherence control in the present *controlled echo*. Second, the first control pulse B whose pulse area is $\pi$ coherently transfers $\rho_{33}$ into an auxiliary ground state $|2\rangle$ at $t = t_B$ without affecting $\rho_{11}$, resulting in spin coherence excitation $\rho_{12}$ (see Fig. 1(c) and the left inset in Fig. 1(d)). Here, the essence of CCC is in the phase relationship between $\rho_{13}$ and $\rho_{12}$ during the quantum coherence control process (discussed below; see *Appendix A*). Third, the newly excited (or coherently transferred) spin coherence $\rho_{12}$ by the first control pulse B starts to diphase according to the spin inhomogeneity much faster than the spin phase relaxation process. This spin dephasing phenomenon is unique in solids but absent in atomic media unless a magnetic gradient field is intentionally applied. Fourth, the Raman rephasing pulse R at $t = t_R$ swaps populations between two ground states $\left(\rho_{11} \xleftrightarrow{R(2\pi)} \rho_{22}\right)$ and results in a Raman echo at $t = t_C$, where $t_C = 2t_R - t_D$ (see the right inset in Fig. 1(d)). The Raman rephasing can be replaced by spin rephasing through the use of a direct rf/microwave π pulse, sacrificing all-optical processing [24]. Finally, the Raman echo is coherently read-out by the control pulse C at $t = t_C$ via transferring $\rho_{22}$ back into the excited state $|3\rangle$, resulting in an optical coherence $\rho_{23}$ at $t = t_e$ (see the red curve in Fig. 1(c)). Here the control pulse access is for the transition of the input pulse A, which is counter-intuitive. This quantum coherently controlled optical coherence $\rho_{23}(t_e)$ is the inverted coherence of $\rho_{13}(t_A)$ and behaves as a macroscopic coherence burst like photon echoes. Thus, the optical photon(s) **e** is emitted via the NDFWM process among A, B, and C. Moreover, the signal **e** can be as high as A in intensity if a backward NDFWM scheme is applied [17,27,41]. Although general physics is the same as conventional quantum nonlinear optics, the counter-intuitive coherence control access of C is new, where the generated NDFWM signal e is like a storage-time extended photon echo. Thus, the present *controlled echo* explores a new realm of ensemble-based quantum memories.

For a perfect population transfer in a lambda-type ensemble, a counter-intuitive pulse sequence between A and B has already been intensively studied in the name of adiabatic passage [45]. Due to a specific pulse shape with long pulse duration in the adiabatic passage, however, the consecutive pulse sequence in Fig. 1(b) may be preferable for mass information processing, sacrificing some coherence loss. Although the coherence loss caused by imperfect population transfer degrades retrieval efficiency, it does not deteriorate quantum fidelity, unless spontaneous emission decay is influenced. By the way, for the resonant Raman excitation in Fig. 2, however, the population shelving (or coherence loss) in the excited state $|3\rangle$ reaches up to 50%, and so does the retrieval efficiency. For a perfect rephasing via complete population swapping between two ground states $|1\rangle$ and $|2\rangle$, the pulse area $\Phi_R$ $\left(\Omega_R = \sqrt{\Omega_A^2 + \Omega_B^2}\right)$ of the Raman rephasing pulse R in Figs. 1 and 2 must be $2\pi$ and balance each other [44]: $\Omega_A = \Omega_B$. Here, the definition of the pulse area $\Phi_i$ is $\Phi_i = \int_0^{\Delta T} \Omega_i dt$, where $\Omega_i$ is the Rabi frequency of pulse *i*. Each optical pulse duration is set to be 0.1 μs for the present numerical calculations. For data pulse A, the pulse area is set to be weak ($\Phi_A = \pi/10$), where $\Phi_B = \Phi_C = \pi$. The magnitude of $\Phi_A$ does not affect the density matrix physics in the *controlled echo* (see the Supplemental Material Fig. S1 for the comparison between $\Phi_A = \pi/2$ and $\Phi_A = \pi/100$). All decay rates are set to zero except for optical phase (homogeneous) relaxation rates $\gamma_{ij}$. Rare-earth doped solids are persistent spectral hole-burning media [42], so that the given optical inhomogeneous width (>GHz) can be tailored to be narrower by the laser jitter (~MHz) of a commercial laser system [2,5,6,9,26,27]. A simple analytical treatment is given for an accelerated optical homogeneous width affected by the optical inhomogeneity as proven experimentally [27]: kHz → MHz. If there is a pulse delay between A and B, the optical phase evolution must be considered together with the spin phase evolution, resulting in overall decoherence (discussed elsewhere).

The novelty of the present *controlled echo* is in the counter-intuitive access of the control pulse C with respect to conventional quantum nonlinear optics [4-14]. In spin homogeneous media such as alkali atoms [4,7,8,10,17,33,38], the control pulse C must be identical to B for the transition $|2\rangle - |3\rangle$ to form a pair pulse via CCC [7,8,10,17], resulting in **e'** at the same frequency as A. In a spin inhomogeneous medium, however, the control pulse access needs to be the opposite for the transition $|1\rangle - |3\rangle$ to avoid population inversion caused by the Raman rephasing between states $|1\rangle$ and $|2\rangle$ (see Fig. 1(d)). The Raman (or spin) rephasing is an essential step toward the *controlled echo*, otherwise the retrieved signal **e** degrades exponentially as a function of C-delay due to spin inhomogeneity (see the Supplemental Material Fig. S2) [20,27]. Even though such opposite control pulse access has been demonstrated in both cw [4,5] and (non-delayed) pulsed [26] schemes without spin rephasing, the retrieved signal is under population inversion. Here, the main check point of the counter-intuitive control pulse access in Fig. 1 is whether the coherence $\rho_{23}(t_e)$ of the echo signal **e** is emissive with or without population inversion. To answer this question, first, the density matrix $\rho_{ab}$ in the interaction Hamiltonian H is analytically investigated:

$$\dot{\rho}_{ab} = -\frac{i}{\hbar}[H, \rho_{ab}], \qquad (1)$$

where $H = -\hbar\Omega$ and $\rho_{ab} = \rho_{ba}^*$.

Before discussion of the *controlled echo* in Fig. 1(a), let us first analyze the conventional case with C applied for the transition $|2\rangle - |3\rangle$ without spin (Raman) rephasing. According to the CCC theory [18,19], the control pulses B and C together result in controlled Rabi flopping between states $|2\rangle$ and $|3\rangle$ and induce coherence inversion to the original $\rho_{13}(t_A)$ excited by the data pulse A (see *Appendix A*; see also the Supplemental Material Fig. S3):

$$\rho_{13}(t_e) = -\rho_{13}(t_A), \qquad (2)$$

where $\rho_{12}(t_B) = -i\rho_{13}(t_A)$. $\qquad (3)$

It should be noted that the density matrix ρ(t) satisfies unitary evolutions which are reversible in the time domain, satisfying quantum memory. On the contrary, the function of a π-rephasing pulse ($R_\pi$) in conventional photon echoes is to invert the absorptive coherence of a data pulse, resulting in an emissive echo: $\rho \xrightarrow{R_\pi} \rho^*$. Here, there is no sign change in dispersion (the real part of coherence $Re\rho$) due to symmetric relation (see Appendix B; see also the Supplemental Material Fig. S4). Thus, the most important check point in Eq. (2) for CCC is whether the same emissive coherence condition of photon echoes is satisfied. The answer is yes, because the sum of the real components of ρ is always zero due to its symmetric distribution (see Fig. 1(f); also see the Supplemental Material Fig. S5) [46]. As experimentally demonstrated with the conventional control pulse access without rephasing effects [7,8], Eq. (2) supports the quantum memory condition in a lambda-type ensemble *iff* the spin transition is homogeneous.

Now, we drive an analytical solution of the *controlled echo* shown in Fig. 1(a) for the counter-intuitive control pulse C resonant between states $|1\rangle$ and $|3\rangle$. Here it should be noted that obtaining an analytic solution directly from Eq. (1) is not possible unless stead-state conditions are given. This is the reason why the numerical calculations are necessary for a pulsed scheme of Fig. 1. However, by using the CCC theory of Eq. (2) and the rephasing condition of $\rho_{21}(t_C) = \rho_{21}^*(t_B)$, the coherence $\rho_{23}(t_e)$ of the *controlled echo* **e** in Fig. 1(a) can be expressed by:

$$\rho_{23}(t_e) = -i\rho_{21}(t_C), \qquad (4)$$

where $\rho_{21}(t_C) = \rho_{12}^*(t_C)$. Because $\rho_{12}^*(t_C) = \rho_{12}(t_B)$ due to the Raman rephrasing, the analytic solution of Eq. (1) for Fig. 1(a) is obtained as:

$$\rho_{23}(t_e) = -\rho_{13}(t_A). \qquad (5)$$

Equation (5) is the same as the conventional one in Eq. (2), resulting in coherence inversion. As mentioned above, Eq. (5) has no spontaneous emission noise due to R-induced population swapping. Thus, the present *controlled echo* scheme of Fig. 1(a) supports the quantum memory with emissive coherence under no population inversion via unitary transformation of Raman (spin) rephasing.

According to the theory [3,17,41,47] and experiments [7-9,20-25,29-33], the density matrix ρ does not discriminate whether the system interacts with a single photon (quantum) or many photons (classical) (see the Supplemental Material Fig. S1). The numerical results will be discussed in Figs. 1(c) and (d).

For the purpose of comparison, a conventional control pulse access in Fig. 1(e) is analyzed as a wrong reference, where the control pulse C is for the $|2\rangle - |3\rangle$ transition. By using the CCC theory of Eq. (2) and the rephasing condition of $\rho_{12}(t_C) = \rho_{12}^*(t_B)$, the coherence $\rho_{13}(t_{e'})$ of the signal **e'** in Fig. 1(e) can be expressed by:

$$\rho_{13}(t_{e'}) = -i\rho_{12}^*(t_C). \quad (6)$$

Thus, by substituting Eq. (3) the analytic solution of **e'** is obtained:

$$\rho_{13}(t_{e'}) = \rho_{13}^*(t_A). \quad (7)$$

Here, Eq. (7) represents rephrased coherence like a photon echo, where the resultant coherence $\rho_{13}$ at t=$t_{e'}$ is also emissive (positive) like Eqs. (2) and (5). Thus, the conventional control access of Eq. (2) applied to a spin inhomogeneous optical ensemble with Raman (spin) rephasing also results in the emissive coherence as experimentally demonstrated in ref. 9. However, $\rho_{13}(t_{e'})$ for **e'** in Eq. (7) is under population inversion due to the Raman rephasing at t=$t_R$. Thus, the conventional control access in a spin inhomogeneous ensemble [9] should fail for quantum memories (numerically discussed in Fig. 1(e)). Therefore, the quantum coherence control in the ensemble media of Fig. 1(a) always results in an emissive NDFWM signal **e** (**e'**) regardless of the control pulse access for either the $|1\rangle - |3\rangle$ or $|2\rangle - |3\rangle$ transition. However, the conventional control pulse access for the $|2\rangle - |3\rangle$ transition must be avoided in a spin inhomogeneous ensemble due to population inversion-caused quantum noises [9,27,44] or absorptive coherence [20] if quantum information processing is targeted.

If there is no Raman rephasing for the conventional control access, the solution of **e'** signal in Eq. (7) is rewritten as:

$$\rho_{13}(t_{e'}) = -i\rho_{12}(t_C) = -\rho_{13}(t_A)e^{-(t_C-t_B)/T_2^*}. \quad (8)$$

where the decay constant $1/T_2^*$ is the spin inhomogeneous width. As expected, Eq. (8) is the same as Eq. (2) except for the decay parameter. Thus, the pulse delay of C from B definitely results in decoherence as experimentally demonstrated [20,27]. This is a clear distinction between spin homogeneous media [7,8,17] and spin inhomogeneous media [9,27] for quantum coherence control. Here, it is not too difficult to expect that Eq. (8) turns out to be absorptive if optical rephasing is added (see the Supplemental Material Fig. S5). To fix this absorptive problem in **e'** an additional $2\pi$ control Rabi flopping of Eq. (2) is required [18,19,27,44,46]. Reference 20 based on Eq. (8) with optical rephasing also need additional $2\pi$ control Rabi flopping to make AFC echo emissive. In that sense, the conventional control pulse access in refs. 9 and 20 is definitely wrong for the quantum memory applications, even though it can still be applicable at low retrieval efficiency [20,48] or beneficial for classical applications [9] such as associative optical memories [49].

Figures 1(c)-(f) show the numerical calculation results of Figs. 1(a) and (b) for two different accesses of the control pulse C: see *Methods* for details of the numerical calculations. Figure 1(c) represents the results of the present *controlled echo* in Figs. 1(a) and (b), where the control pulse access is counter-intuitive with respect to conventional ones (see Fig. 1(e)). In Fig. 1(c), the data pulse-excited coherence $\rho_{13}(t_A)$ (see the inset) is transferred into spin coherence $\rho_{12}$ (= $-i\rho_{13}$; see the green curve) by the first control pulse B in a Raman scheme as discussed in Eq. (3). The dotted curve is for the excited state population $\rho_{33}$, where its evolution directly influences the optical-spin

coherence conversion. As discussed above analytically, the Raman rephasing pulse R swaps the population and rephrases the spin coherence (see the green curve): $\rho_{12} \xrightarrow{R(2\pi)} \rho_{12}^*$. The rephased coherence ($\rho_{12}^*$) is coherently read out by C, resulting in the emissive NDFWM signal **e** [$Im\rho_{23}(t_e)$] (see the red curve), as discussed in Eq. (5).

To support Figs. 1(c)~(f) ensemble phase evolutions are introduced and analyzed for the *controlled echo*. The coherence evolution of the present *controlled echo* sequence can be simply described by:

$$\rho_{12}(t) = -i\rho_{13}(t_A)e^{\pm i\delta_j t}: t_B \leq t < t_R, \tag{9-1}$$

where $\delta_j$ is the detuning of the j$^{th}$ spin. By the Raman rephasing pulse R at t=$t_R$, the spin coherence of Eq. (9-1) is rephased:

$$\rho_{12}(t') = i\rho_{13}^*(t_A)e^{\pm i\delta_j(t'-T)}: t_R \leq t' < t_C, \tag{9-2}$$

where $T \equiv (t_R - t_B) = (t_C - t_R)$ and $t' = t - T$. Therefore,

$$\rho_{12}(t_C) = i\rho_{13}^*(t_A), \tag{9-3}$$

where $t_C = t_B + 2T$. For the quantum coherence control by the second control pulse C whose transition is for $|2\rangle - |3\rangle$ (conventional access),

$$\rho_{13}(t'') = -i\rho_{12}(t_C)e^{-\frac{t''}{T_2}} = \rho_{13}^*(t_A)e^{-\frac{t''}{T_2}}: \ t'' \geq t_C, \tag{9-4}$$

where $t'' = t - t_C$. On the contrary, for the quantum coherence control by the second control pulse C, whose transition is for $|1\rangle - |3\rangle$,

$$\rho_{23}(t'') = -i\rho_{21}(t_C)e^{-\frac{t''}{T_2}} = -\rho_{13}(t_A)e^{-t''/T_2}. \tag{9-5}$$

Thus, both NDFWM signal **e** and **e'** represent the same emissive coherence. Because all real components are zeros at $t = t_e$, Eq. (9-5) is the same as Eq. (9-4). Since Eq. (9-5) is free from the spontaneous emission noise due to no population inversion, however, only the counter-intuitive control pulse access is accepted as a quantum memory protocol:

$$\rho_{23}(t_e) = -\rho_{13}(t_A). \tag{10}$$

The magnitude of **e** $(Im\rho_{23}(t \geq t_e))$ in Fig. 1(c) decays down exponentially as a function of time by the predetermined optical phase relaxation rate $\gamma_{23}$ (1/$T_2$) as denoted by Eq. (9-5). This decay is of course accelerated by optical inhomogeneous broadening. The excited state population $\rho_{33}$ (see the dotted line) for **e** at t=$t_e$ is exactly the same as that for A at t=$t_A$, as shown in the inset (see also Fig. 2(d)), resulting in no spontaneous emission-caused quantum noise. Figure 1(d) shows population evolutions for Fig. 1(c), where the data A-excited population $\rho_{33}$ is fully transferred into state $|2\rangle$ by B (see the left inset). As shown in the right inset, the balanced Raman rephasing pulse R at t=$t_R$ swaps the population between two ground states $|1\rangle$ and $|2\rangle$: $\rho_{11} \xleftrightarrow{R(2\pi)} \rho_{22}$. Thus, the present *controlled echo* in a lambda-type three-level solid ensemble whose spin transition is inhomogeneous is satisfied for a quantum memory protocol, as discussed with the analytical proofs of Eqs. (5) and (10), and as numerically demonstrated in Fig. 1(c). Here, it should be noted that the higher Raman Rabi frequency weakens the shelved atom-based coherence loss. This is the only reason why much stronger R is used in the calculations, even though it does not affect too much in real media.

In contrast to Fig. 1(c), Fig. 1(e) is for the conventional control pulse access for the $|2\rangle - |3\rangle$ transition, also resulting in an emissive **e'** signal at the same frequency as A for the $|1\rangle - |3\rangle$ transition as shown in the analytical expression of Eqs. (7) and (9-4). When $t < t_C$, all coherence evolutions are exactly the same as Fig. 1(c). However, the emissive NDFWM signal **e'** at $t \geq t_e$ is under population inversion ($\rho_{33} \gg \rho_{11}$) due to the Raman pulse-induced population swapping between two ground states $|1\rangle$ and $|2\rangle$. Thus, Fig. 1 supports the present counter-intuitive control pulse access for the *controlled echo*-based quantum memory and provides general physics of quantum coherence control in a spin inhomogeneously broadened solid ensemble.

The symmetric coherence evolutions appeared in the real parts of coherence in the retrieved optical signal **e** ($\rho_{23}$) and **e'** ($\rho_{13}$) analyzed in Eqs. (5) and (7) are numerically demonstrated for different control pulse access in Fig. 1(f). Due to the symmetric distribution of the real parts in optical coherence in both $Re\rho_{13}$ and $Re\rho_{23}$, the overall sum-coherence is zero all the time in both cases (see the Supplemental Material Fig. S4). If there is no Raman rephasing, quantum memory cannot be satisfied due to spin dephasing, resulting in zero conversion efficiency for **e** (**e'**). In spin homogeneous media like alkali atoms [7,8,33], however, the Raman rephasing must be prevented, and the retrieval efficiency is inversely proportional to the spin phase relaxation time. Thus, the spin (Raman) rephasing plays a crucial role in a spin inhomogeneous solid ensemble not only for storage time extension but also for fidelity enhancement. This is the basic physics of quantum coherence control in an optical ensemble analyzed with exact coherence evolutions, and explains how the blind adaptation of conventional CCC demonstrated in atomic media has misled the quantum optics community [9,20], where ref. [9] is due to population inversion, and ref. [20] is due to absorptive coherence for echo signals.

Figure 2 is for a resonant Raman data pulse for Fig. 1, where the optical pulses A and B form a simultaneous Raman data pulse D [6,44]. The pulse area of A is kept the same as Fig. 1. For maximum coherence excitation, the Raman data pulse area is $\Phi_D=\pi$, where $\Omega_D = \sqrt{\Omega_A^2 + \Omega_B^2}$ and $\Omega_A \ll \Omega_B$. All others are the same as in Fig. 1. As shown in Fig. 2(a), the present protocol of *controlled echo* is still working for the resonant Raman data for quantum memories. However, the excited population by the data pulse A is half-shelved in the excited state $|3\rangle$, where $\rho_{22}=\rho_{33}$ (see Fig. 2(b)). Because the coherence is induced by the population difference, e.g., $\rho_{12} \propto \rho_{22} - \rho_{11}$, the spin coherence reduction in the resonant Raman case is severe up to 2/3 for $\Omega_A \ll \Omega_B$ and $\Phi_A \ll 1$ (see Figs. 2(c) and (d)). In Fig. 2(a), the reduced coherence on $Im\rho_{13}$ (and thus $Re\rho_{12}$) is ~67% in magnitude. Moreover, the shelved population on $|3\rangle$ may deteriorate the read-out conversion efficiency by C. The shelved atom-caused coherence loss should deteriorate both the quantum fidelity by spontaneous emission decay and retrieval efficiency.

The gradient Raman echo based on off-resonant interactions may be free from the population shelving or population decay-caused quantum noises discussed in Fig. 2 [33]. Instead of the direct Raman rephasing in Fig. 2, an oppositely polarized gradient control pulse set functions the Raman rephasing in ref. 33. In that sense, ref. 33 is in the same category as the present quantum coherence control. However, with the use of a common control pulse set (B and C) for multiple data pulses (As) for the Raman gradient effect, 50% coherence loss at least is inevitable due to 'ON' of the control pulse C for 'OFF' data pulse excited Raman coherence timing. Moreover, such a Raman gradient technique can never be applied to a solid ensemble due to the intrinsic spin inhomogeneity. Although the gradient technique in both optical [22] and Raman [33] schemes is best for a forward propagation scheme to overcome the intrinsic low retrieval efficiency *via* circumventing the echo resorption problem in an ensemble, simultaneous satisfaction of both ultralong storage and near perfect retrieval efficiency is impossible in an optical regime.

Figure 3 is for the present *wavelength convertible quantum memory* in a double lambda-type four-level optical medium whose spin transition is inhomogeneouly broadened. As shown in Fig. 3(a), an extra state $|4\rangle$ is simply added in Fig. 1(a), where state $|4\rangle$ is used for both independent Raman rephasing and wavelength conversion. Unlike Fig. 1, the resonant Raman rephasing pulse R composed of light pulses C and D is applied for the transition $|1\rangle - |4\rangle - |2\rangle$ (see Fig. 3(a)), so that any potential defect by the shelved atoms on state $|3\rangle$ can be removed on the coherence recovery process. Here, a frequency difference between A and **e** is the origin of the frequency up- or down-conversion. The control pulse (Cn) access is for the transition $|1\rangle - |4\rangle$ and results in the photon signal **e**, resonant between states $|2\rangle$ and $|4\rangle$ via both spin-optical coherence conversion and Raman rephasing. Figure 3(b) is the pulse sequence of Fig. 3(a), where the scale of the Raman rephasing R (C and D) is intentionally cut down to show the same consecutive pulse sequence of A and B as in Fig. 1. The Rabi frequency $\Omega_A$ of the data pulse A in Fig. 3 is decreased by a factor of $\sqrt{2}$ for the purpose of comparison with Fig. 1. As a direct result, the coherence excitation by A is also decreased by $\sqrt{2}$ in Fig. 3(c) (see also the inset of Fig. 3(d)). Either for $|3\rangle$ or $|4\rangle$, a resonant Raman pulse deals only with two-photon coherence between the two ground states $|1\rangle$ and $|2\rangle$, resulting in the same result but different frequency (see Figs. 1(c) and 3(c)).

The retrieved photon signal **e** must satisfy Kerr nonlinear optics among pulses A, B, and Cn, resulting in up- or down-conversion depending on the relative energy level of $|2\rangle$. The control pulse Cn can be applied for either transition $|1\rangle - |4\rangle$ as in Fig. 3(c) or $|2\rangle - |4\rangle$ as in Fig. 3(d). Such a double lambda system has never been studied for quantum memories based on coherent transients. Neither the ensemble phase control nor the coherent transients have been discussed ever. So far, general understanding of the physical mechanism of Kerr nonlinear optics with coherent transients has never been achieved for quantum memories.

Figure 3(c) shows the numerical results of Figs. 3(a) and (b), where all decay rates are set to zero except for the optical phase relaxation rates. For the purpose of simplification, the optical homogeneous decay rate is replaced by laser jitter-dependent optical inhomogeneous width in a persistent spectral hole-burning medium as explained in Fig. 1. As shown in Fig. 3(c), the NDFWM signal **e** ($Im\rho_{24}$; red) as a coherence retrieval of data A via Raman rephasing shows the same emissive coherence as in Eq. (5) for the counter-intuitive control pulse access of Cn. The dotted line represents the excited state population $\rho_{44}$, where $\rho_{44}$ at t=$t_e$ is the same as the A pulse-excited $\rho_{33}$ at t=$t_A$ (see the inset of Fig. 3(d)). Thus, Fig. 3(a) is free from the spontaneous emission noise problem. The counter-intuitive control pulse access is of course to compensate the population swapping between two ground states by the resonant Raman rephasing pulse R composed of C and D to satisfy no population inversion for the signal **e**: $\rho_{11} \xleftrightarrow{R(2\pi)} \rho_{22}$.

In contrast to Fig. 3(c), the control pulse (Cn) access is applied for the transition $|2\rangle - |4\rangle$ in Fig. 3(d), which is conventional as in Fig. 1(e). The resultant coherence of the NDFWM signal **e'** denoted for $Im\rho_{14}$ is also emissive, but under population inversion ($\rho_{44}$~1 at t=$t_{e'}$). Even if such an inverted echo signal may be useful for classical applications of associative memories [49], the potential quantum noise from $\rho_{44}$ should prevent Fig. 3(d) from quantum memories. As discussed in Fig. 1(f), the coherence inversion in Eq. (5) is also shown in Figs. 3(e) and (f) for the real parts of the signals **e** and **e'** for the opposite control pulse access in Figs. 3(c) and (d), respectively. This proves again a clear distinction between rephasing-based quantum memories and the present *controlled echo*: $\rho^*$ vs. $-\rho$. As mentioned in Fig. 1, the sum of $Re\rho_{13}$ (or $Re\rho_{24}$) is always zero due to its symmetric distribution, so that the coherence inversion effect in Figs. 3(c) and (d) is like rephasing of photon echoes, where the NDFWM signal (**e** or **e'**) always satisfies the phase matching condition and emissive coherence regardless of the control pulse access either for $|2\rangle - |4\rangle$ or $|1\rangle - |4\rangle$ transition. The only matter is whether the echo signal **e** is free from the spontaneous emission-caused quantum noise or not. Thus, the present *wavelength convertible quantum memory* protocol in a four-level system is proven for the wavelength conversion, which is essential for future spectral division multiplexing in quantum networks.

In general, the maximum spin coherence $\rho_{12}$ indirectly excited by a consecutive optical pulse set A and B exponentially decreases as a function of the pulse delay ΔT between them, where the loss factor is $e^{-\eta}$: $\eta = \Delta T \left( \Delta_{inh}^{opt} + \frac{1}{T_2^{opt}} \right)$. For $\Delta_{inh}^{opt} \gg \gamma_{opt} (= T_2^{opt})$, the optical inhomogeneous broadening becomes a dominant factor on the coherence loss. Thus, as done in all calculations of this research, the replacement of the optical homogeneous decay rate by the laser jitter-modified optical inhomogeneous broadening is reasonable [27]. With a zero pulse delay as in Figs. 1 and 3, the coherence loss is nearly neglected even in an inhomogeneously broadened optical ensemble. The spin inhomogeneity may also affect the coherence loss due to different detuning effects on the different optical transitions of either $|1\rangle - |3\rangle$ or $|2\rangle - |3\rangle$. As shown in the inset of Fig. 3(d), the optical-spin coherence conversion by the first control pulse B is not perfect due only to this reason: $|Re\rho_{12}| < |Im\rho_{13}|$. In an actual medium, e.g., $Pr^{3+}$-doped $Y_2SiO_5$, however, this spin inhomogeneity effect is negligibly small because the optical Rabi frequency (~MHz) of B is much greater than the spin inhomogeneous broadening (30 kHz). In a persistent spectral hole-burning medium, the minimum optical inhomogeneous width of 1.6 MHz for the 0.1 μs pulse duration in Fig. 3 can be easily obtained by using a commercial ring-dye laser system and acousto-optic modulators [2,5,6,9,27,26]. Once again the stronger Raman pulse R in Figs. 1~3 is not necessary in a real system.

For the near perfect retrieval efficiency in the present *controlled echo* quantum memory, the control pulse propagation directions must be backward each other, so that a backward echo signal **e** can be generated by the following phase matching conditions:

$$\omega_e = -\omega_A + \omega_B + \omega_{Cn}, \tag{11-1}$$

$$\boldsymbol{k}_e = -\boldsymbol{k}_A + \boldsymbol{k}_B + \boldsymbol{k}_{Cn}, \tag{11-2}$$

where $\omega_j$ ($\boldsymbol{k}_j$) is the angular frequency (wave vector) of pulse *j*, and the subscript 'e' stands for the *controlled echo*. Although a perfect collinear scheme between A and **e** cannot be satisfied due to $\boldsymbol{k}_B + \boldsymbol{k}_{Cn} \neq 0$, the wavelength deviation among them is negligibly small at $10^{-8}$ for most rare-earth doped solids [42]. Thus, the refractive index-dependent phase walk-off is also negligibly small at far less than π as experimentally demonstrated [5,6,9,26,27]. This flexible NDFWM offers a great advantage in photon echo-based quantum memories for spatial multiplexing. According to the theory [17] and experimental observations [27], a near perfect retrieval efficiency $\eta_e$ can be obtained for the backward controlled echo scheme even for an optically thick ensemble: $\eta_e = \left( 1 - e^{-\alpha l} \right)^2$ [41]. Here, the higher optical depth (*αl*) is necessary for the single write- and read-out process, otherwise an optical cavity is needed sacrificing bandwidth. The phase matching conditions of Eqs. (10) and (11) can also be expanded for polarizations.

In conclusion, a *wavelength convertible quantum memory* protocol based on *controlled echo* was introduced, analyzed, and discussed for a spin inhomongeneously broadened double lambda-type four-level optical ensemble. Unlike alkali atoms whose spin transitions are homogeneous, the control pulse access in a spin inhomogeneous solid ensemble must be counter-intuitive to avoid spontaneous emission-caused quantum noise. In the present study, quantum coherence control in a spin inhomogeneously broadened solid ensemble was explicitly discussed to elucidate the basic but novel physics of ensemble phase control for quantum coherence conversion and Raman rephasing, resulting in near perfect, ultralong, and emissive photon echoes without quantum noises. With a backward control pulse and balanced Raman rephasing, the retrieval efficiency can be near perfect due to the absence of echo reabsorption, and the photon storage time can be extended up to the spin homogeneous decay time. Here, the spin homogeneous decay time can be as long as minutes or even hours in rare-earth doped solids [37]. Moreover, the present controlled echo scheme is simple in configuration and applicable to spectral and spatial multiplexing in all-optical quantum information processing in the future quantum networks. The present research sheds light on potential quantum memory applications in various quantum information areas such as scalable qubit generations, recursive operations, sensing, and quantum repeaters for long-distance quantum communications.

*Methods:* For the numerical calculations, total sixteen time-dependent density matrix equations are solved for a four-level ensemble medium in an interacting Heisenberg picture under rotating wave approximations [47]: $\frac{d\rho}{dt} = \frac{i}{\hbar}[H,\rho] - \frac{1}{2}\{\gamma,\rho\}$, where $\rho$ is a density matrix element, H is Hamiltonian, and $\gamma$ is a decay parameter. In the calculations, 99.55% of the Gaussian distribution is taken for total 201 distributed spin groups at 2 kHz spacing, where the spectral spin inhomogeneous width (FWHM) is set at 170 kHz. Here, the exaggerated spin bandwidth (x30) is only to save computer calculation time. Those 201 spectral groups are calculated for the time domain and summed up for all spectral groups. For the optical transition, optical inhomogeneity is neglected because it does not violate physics of Raman coherence nor affect the result, unless optical pulse delay between A and B is given. The following equations are for the coherence terms of $\dot{\rho}_{ij}$ in a four-level system interacting with three resonant optical fields:

$$\frac{d\rho_{12}}{dt} = \frac{i}{2}[\Omega_1\rho_{32} - \Omega_2\rho_{13} + \Omega_3\rho_{42} - \Omega_4\rho_{14}] - i\delta_{12}\rho_{12} - i(\delta_2-\delta_1)\rho_{12}, \qquad (12\text{-}1)$$

$$\frac{d\rho_{13}}{dt} = \frac{i}{2}[\Omega_1(\rho_{33} - \rho_{11}) - \Omega_2\rho_{12}] - i\delta_1\rho_{13} - \gamma_{13}\rho_{13}, \qquad (12\text{-}2)$$

$$\frac{d\rho_{14}}{dt} = \frac{i}{2}[\Omega_3(\rho_{44} - \rho_{11}) - \Omega_4\rho_{12}] - i\delta_3\rho_{14} - \gamma_{14}\rho_{14}, \qquad (12\text{-}3)$$

$$\frac{d\rho_{23}}{dt} = \frac{i}{2}[\Omega_2(\rho_{33} - \rho_{22}) - \Omega_1\rho_{21}] - i\delta_2\rho_{23} - \gamma_{23}\rho_{23}, \qquad (12\text{-}4)$$

$$\frac{d\rho_{24}}{dt} = \frac{i}{2}[\Omega_4(\rho_{44} - \rho_{22}) - \Omega_3\rho_{21}] - i\delta_4\rho_{24} - \gamma_{24}\rho_{24}, \qquad (12\text{-}5)$$

$$\frac{d\rho_{44}}{dt} = \frac{i}{2}[\Omega_3(\rho_{14} - \rho_{41}) + \Omega_4(\rho_{24} - \rho_{42})] - (\Gamma_{41} + \Gamma_{42})\rho_{44}, \qquad (12\text{-}6)$$

where the interaction Hamiltonian matrix H is given by:

$$H = -\frac{\hbar}{2}\begin{bmatrix} -2\delta_1 & 0 & \Omega_1 & \Omega_3 \\ 0 & -2\delta_2 & \Omega_2 & 0 \\ \Omega_1 & \Omega_2 & -2\delta_3 & 0 \\ \Omega_3 & 0 & 0 & 0 \end{bmatrix}. \qquad (13)$$

Here $\Omega_1$ ($\Omega_3$) is the Rabi frequency of the optical field between the ground state $|1\rangle$ and the excited state $|3\rangle$ ($|4\rangle$), and $\Omega_2$ ($\Omega_4$) is the Rabi frequency of the optical field between the ground state $|2\rangle$ and the excited state $|3\rangle$ ($|4\rangle$). The $\delta_1$, $\delta_2$, and $\delta_3$ are the atom detuning from the resonance frequency for $\Omega_1$, $\Omega_2$, and $\Omega_3$ ($\Omega_4$), respectively. For visualization purpose and simplification, all decay terms are neglected except for the optical phase relaxation rates $\gamma_{ij}$.

The optical pulse duration is set at 0.1 μs, otherwise specified. The time increment in the calculations is 0.01 μs. Initially all atoms are in the ground state $|1\rangle$ ($\rho_{11}(0) = 1$), and all initial coherence is $\rho_{ij}(0) = 0$, where $i$ ($j$)=1, 2, 3, 4. The program used for the numerical calculations is time-interval independent, so that there is no accumulated error depending on the time interval settings. For the control (rephaisng) π pulse area, the corresponding Rabi frequency is set at $\frac{100}{\sqrt{2}}$ MHz for the pulse duration of 0.01 μs to satisfy a 2π pulse area. The present numerical calculations of the *controlled echo* have been successfully demonstrated for cw and pulse NDFWM without rephasing, but never been demonstrated for a rephasing scheme.

This work was supported by the ICT R&D program of MSIT/IITP (1711042435: Reliable crypto-system standards and core technology development for secure quantum key distribution network). The author thanks M. O. Scully (Texas A&M University, USA) for helpful discussions in QNO2018 conference held in Malaysia.

**Appendix A:** *Optical-spin coherence conversion*

The uniqueness of CCC is in the phase control of the ensemble coherence via a controlled Rabi flopping of the excited atom(s) [18,19,27,46]. Unlike a π phase shift in a two-level photon echo system, a complete population transfer between the common excited state $|3\rangle$ and an auxiliary ground state $|2\rangle$ in a three-level system induces a π/2 phase shift to the pre-excited optical coherence $\rho_{13}$. Thus, the Rabi flopping by the control pulse set induces a coherence inversion, resulting in an absorptive echo in the controlled rephasing scheme (see the Supplemental Material Fig. S5) [27,46]. In density matrix notations, the π/2 phase shift represents coherence *swapping* between the imaginary and real parts of the coherence: $Re\rho \leftrightarrow Im\rho$. Thus, the function of the control π–pulse of B is expressed as:

$$\rho_{13} \xrightarrow{B(\pi)} \rho_{12} \ (= -i \cdot \rho_{13}). \tag{14}$$

The negative sign is for the consistency of the swapping (see the same value at B for $Im\rho_{13} \rightarrow Re\rho_{12}$ in Figs. 1(c), 2(c), and 3(d)). This is the correct understanding of the physical mechanism for the π–control pulse B in a lambda-type system of Fig. 1(a). Thus, the *coherence inversion* via a complete population oscillation by the control pulse set B and C in Fig. 1 is accomplished:

$$\rho_{13} \xrightarrow{B(\pi)+C(\pi)} -i \cdot (-i \cdot \rho_{13}) = -\rho_{13}. \tag{15}$$

Equations (14) and (15) represent a correct physical mechanism of the conventional quantum coherence control in an ensemble. As discussed, the control Rabi flopping added to AFC echoes in ref. [20] results in an absorptive echo (see the Supplemental Material Fig. S5) [27,46]. As mentioned in the text, the spin inhomogeneity in a solid ensemble causes irreversible decoherence [20,27]. To solve the absorptive echo problem, additional control Rabi flopping must be added [27]. The observation of controlled AFC echoes in ref. [20] is due to imperfect coherence evolutions due to nonuniform Rabi frequency of Gaussian light along the transverse axis, otherwise impossible due to absorptive echo coherence [48].

**Appendix B:** *Photon echo rephasing*

To explain the fundamental difference between π rephasing in two-level photon echoes, and CCC-induced coherence inversion in a three-level sysetem, coherence evolution in photon echoes are presented. For typical light-matter interactions in a two-level system, the relationship between absorption and dispersion is denoted by the Kramers-Kronig (K-K) relation. By the K-K relation one part is obtained by integrating the other part, resulting in sin and cos relationship. The π rephasing in a two-level system is represented by:

$$\text{Symmetry } (Im\rho): \cos(\delta_j t) \xrightarrow{R(\pi)} \cos(\delta_j t + \pi) = -\cos(\delta_j t) = -\cos(-\delta_j t), \tag{16}$$

$$\text{Antisymmetry } (Re\rho): \sin(\delta_j t) \xrightarrow{R(\pi)} \sin(\delta_j t + \pi) = -\sin(\delta_j t) = \sin(-\delta_j t), \tag{17}$$

where $\cos(-\delta_j t)$ and $sin(-\delta_j t)$ represent for a time reversal in coherence evolutions (see the Supplemental Material Fig. S4). Equation (16) represents a symmetric feature, while Eq. (17) is anti-symmetric. Thus, only imaginary part (absorption) of the ensemble coherence ρ has a sign change across the π rephasing pulse as shown Fig. S4(g). The absorption and dispersion relation also indicates that the maximum coherence of the imaginary part (absorption) represents zero in the real part (dispersion). Due to the symmetric distribution of the real parts of coherence (see the Supplemental Material Figs. S4(d) and S4(f)), the sum of real parts in time domain is always zero (see the Supplemental Material Fig. S5(b)). Especially individual real parts are all zero at the photon echo timing.

In a double rephasing scheme of photon echoes the final echo has exactly the same coherence as the data (see the Supplemental Material Figs. S4(i) and (j)) [19]. In other words, the doubly rephased echo cannot be extracted from an optically thick ensemble. In that sense, the modified photon echoes [23,24] based on double rephaisng cannot be applied for quantum memories without CCC [19]. The echo observations in refs. [23,24] are also due to imperfect Rabi frequency in the transverse spatial mode [48].

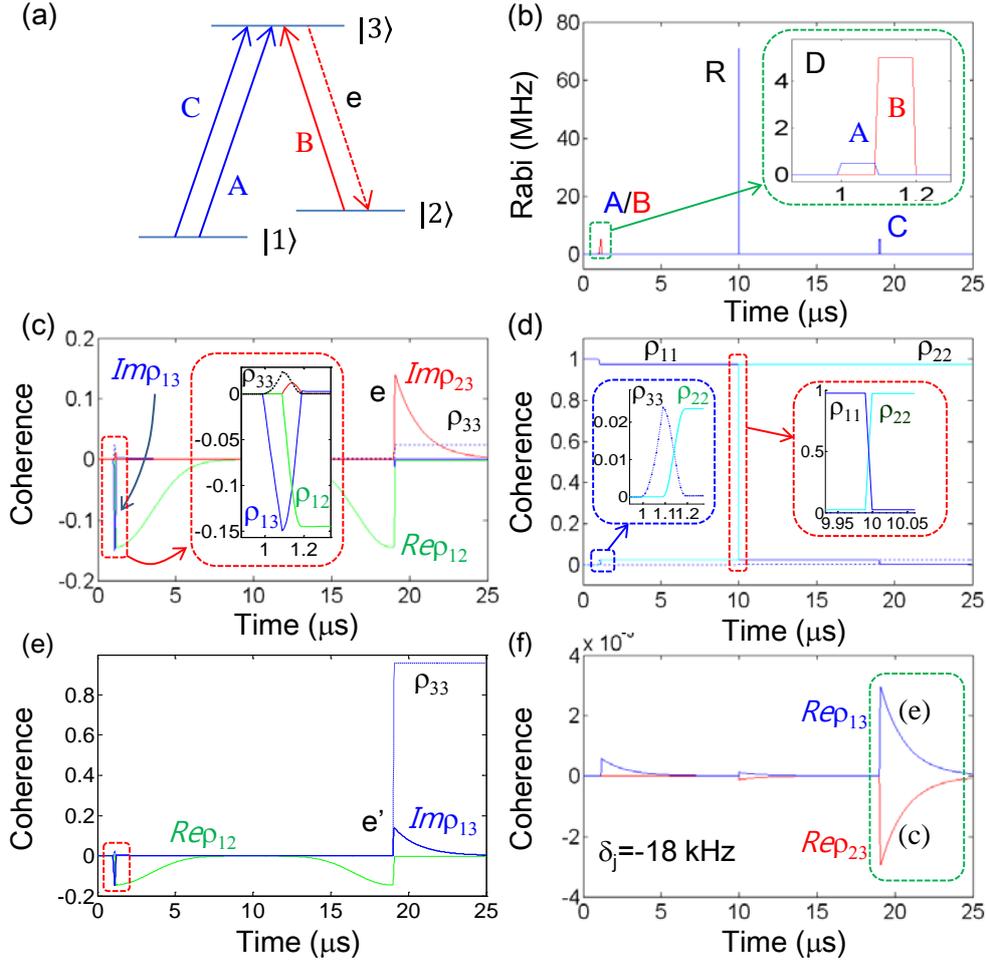

Fig. 1. Controlled echo in a spin inhomogeneously broadened three-level optical ensemble. (a) A lambda-type energy level diagram interacting with optical pulses. (b) Pulse sequence of (a). Resonant Raman rephasing pulse R is composed of equal Rabi frequency of A and B. The control pulse C is resonant for the transition of either $|1\rangle - |3\rangle$ or $|2\rangle - |3\rangle$. The pulse arrival time of A, B, R, and C is $t_A = 1.0$ μs, $t_B = 1.1$ μs, $t_R = 10.0$ μs, and $t_C = 19.0$ μs, respectively. Each pulse duration is 0.1 μs except for R with 0.01 μs. (c) Coherence and (d) Population evolutions of (a) and (b). Blue: $Re\rho_{13}$, Red: $Im\rho_{23}$, Green: $Re\rho_{12}$, Dotted: $\rho_{33}$. (e) Numerical results when C is for $|2\rangle - |3\rangle$ transition: $\rho_{33} \gg \rho_{11}$ at $t > t_{e'}$ (19.1 μs). (f) Coherence evolutions of real components for two different access of C either for (c) or (e). The detuning $\delta_j$ is for the $j^{th}$ detuned spin. All decay rates are zero except for phase relaxation rates $\gamma_{31} = \gamma_{32} = 50$ kHz. The spin inhomogeneous width (FWHM) of $|1\rangle - |2\rangle$ transition is 170 kHz. The retrieved signal **e** (**e′**) is for the transition $|2\rangle - |3\rangle$ ($|1\rangle - |3\rangle$). The Rabi frequency of R is $\Omega_R = 100/\sqrt{2}$ MHz. The Rabi frequency $\Omega_A$, $\Omega_B$, and $\Omega_C$ is 0.5, 5, and 5 MHz, respectively. All numbers in decay rates and Rabi frequencies are divided by $2\pi$.

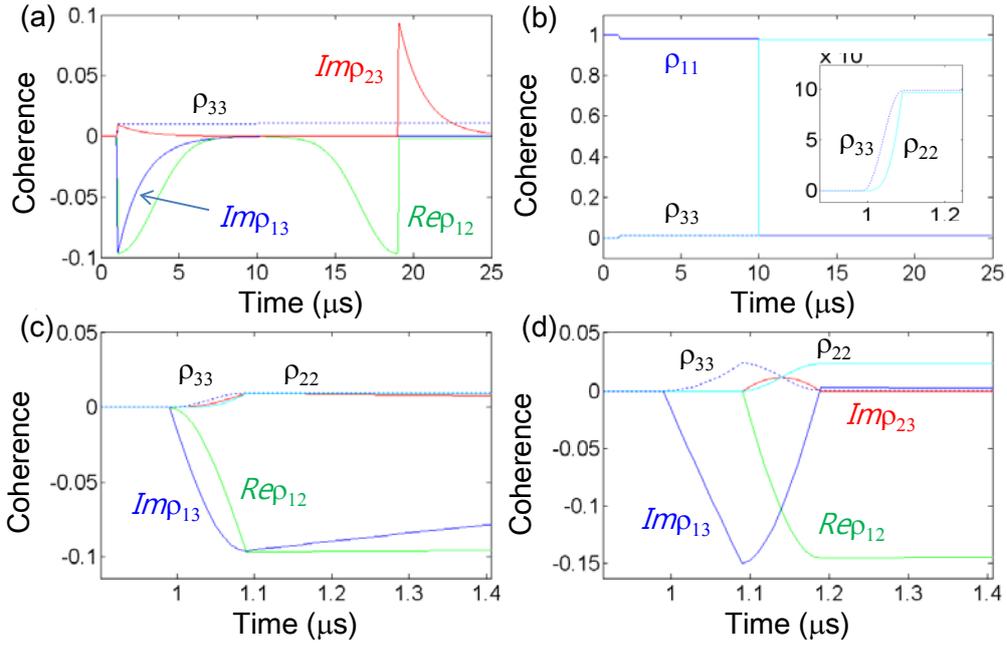

Fig. 2. Controlled echo calculations for resonant Raman data. (a)-(c) Numerical results of Fig. 1(c) when D is composed of a simultaneously resonant Raman pulse set of A and B. (d) Details of Fig. 1(c) corresponding to (c). The pulse area of the data D is π, where the generalized Rabi frequency of D is $\Omega_D = \sqrt{\Omega_A^2 + \Omega_B^2} = 5\ MHz$. All other parameters are the same as in Fig. 1.

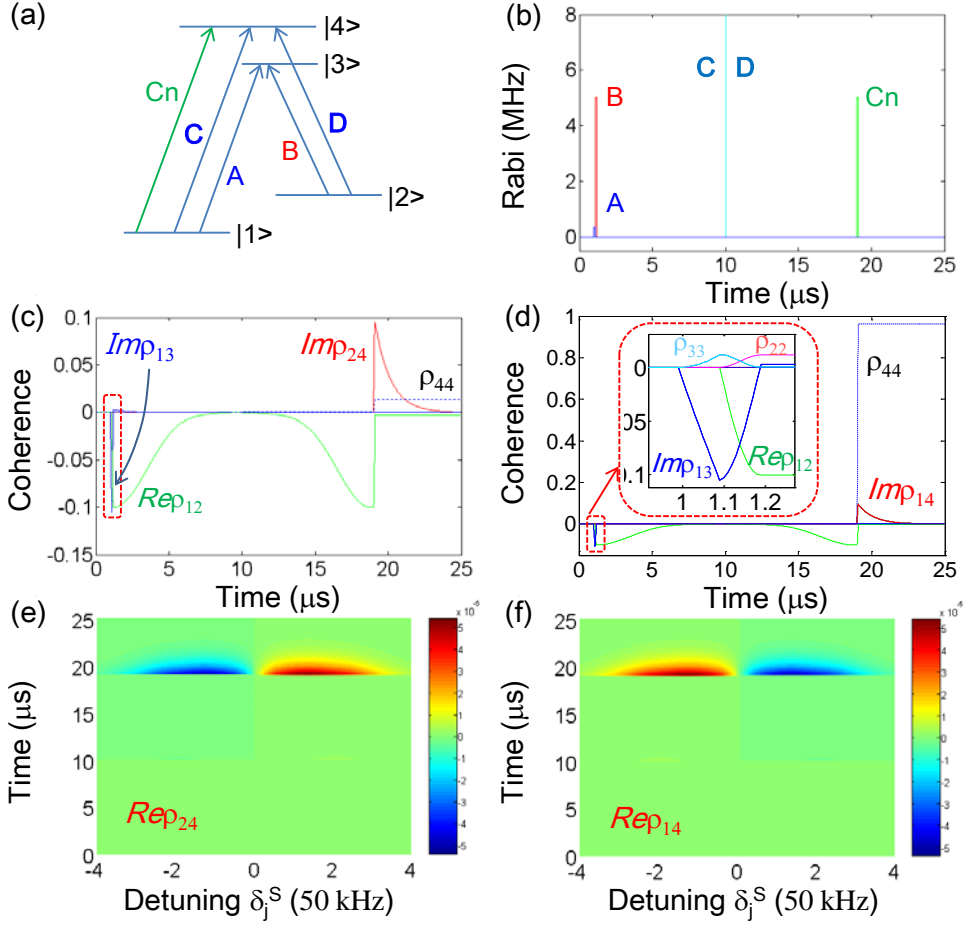

Fig. 3. A wavelength convertible controlled echo. (a) Energy level diagram and (b) pulse sequence. The optical pulses C and D are for resonant Raman rephasing whose pulse area is $2\pi$, and each Rabi frequency is $\Omega_C = \Omega_D = 100/\sqrt{2}$ MHz. The control Rabi frequency Cn is $\Omega_{Cn} = 5$ MHz. (c)-(f) Numerical calculations for (a) and (b), where Cn is for the transition $(|2\rangle - |4\rangle)$ in (d) and (f). The dotted box in (c) is the same as the inset in (d). All decay rates are zero except the optical homogeneous decay rates of 150 kHz. All other parameters are the same as in Fig. 1.